\documentclass[12pt,a4paper,final]{iopart}

\usepackage{iopams}
\usepackage{cite}
\expandafter\let\csname equation*\endcsname\relax
\expandafter\let\csname endequation*\endcsname\relax

\usepackage{amsmath}
\usepackage{iopams}
\usepackage{graphicx}
\usepackage[breaklinks=true,colorlinks=true,linkcolor=blue,urlcolor=blue,citecolor=blue]{hyperref}

\def\P{\mathbb{P}}

\def\S{\mathcal{S}}

\newcommand{\Sup}{\mathrm{sup}}
\newcommand{\sub}{\mathrm{sub}}
\DeclareMathOperator{\cov}{cov}

\begin{document}

\title[]{Path-integrals and optimal paths for the  fractional Ornstein-Uhlenbeck process}

\author{Bing Miao$^{1}$, Gleb Oshanin$^{2}$ and Luca Peliti$^3$ }
\address{$^1$ 
University of Chinese Academy of Sciences (UCAS), Beijing $100049$, China\\
$^2$ Sorbonne Universit\'e, CNRS, Laboratoire de Physique Th\'eorique de la Mati\`{e}re Condens\'ee (UMR CNRS 7600), 4 place Jussieu,
75252 Paris Cedex 05, France\\
$^3$ Santa Marinella Research Institute, I-00058, Santa Marinella, Italy}

\begin{abstract}
	We derive the path-integral representation of the fractional Ornstein-Uhlenbeck process driven by Riemann-Liouville fractional Gaussian noise, for both the subdiffusive and superdiffusive regimes. We express the corresponding action, which is a quadratic functional of individual trajectories of the process, in two alternative but equivalent forms: either as a fractional integral or as a double integral with a nonlocal kernel. Moreover, we determine in closed form the optimal (action-minimizing) paths conditioned to reach a prescribed point at a fixed time moment and discuss their behavior, which appears to be non-intuitive for subdiffusive processes in the presence of a strong confining potential.   
\end{abstract}


\maketitle

\section{Introduction}

The Ornstein–Uhlenbeck process (OUP) was first introduced to describe velocity fluctuations of a Brownian particle under linear damping \cite{ou,doob,wang}. It also models the motion of overdamped particles in optical traps \cite{tweezer} or tethered to polymer backbones \cite{polymer}. As the simplest Gaussian Markov process driven by white noise, the OUP offers a canonical example of mean-reverting relaxation. Its tractability and explicit correlations have made it a standard model in statistical physics and stochastic thermodynamics \cite{pel,pel2,carlos}, with applications extending to finance, evolutionary dynamics, neuroscience, and climate modeling. Recent work \cite{ergo,eli} further analyzes its stochastic properties and highlights a broad scientific relevance of the OUP.

In diverse complex systems such as viscoelastic media, cells, turbulence, finance, and climate, the driving noise often deviates strongly from Gaussian white noise \cite{ralf,target}. 
An extensively studied generalization of the OUP replaces the white-noise term with fractional Gaussian noise (fGn) \cite{man,oks}, which exhibits long-range temporal correlations characterized by the Hurst exponent 
$H$. The resulting fractional Ornstein–Uhlenbeck process (fOUP) is non-Markovian and incorporates memory effects relevant to several forms of experimentally observed anomalous diffusion. Depending on 
$H$, the fOUP displays subdiffusive ($H<1/2$) or superdiffusive ($H>1/2$) behavior, making it a suitable model for correlated stochastic relaxation in diverse physical, biological, and environmental settings. Consequently, its statistical properties and dynamical features have been the subject of substantial analytical and numerical studies (see, e.g., \cite{cheredito,salminen,jeon,chev,bar}).

The analysis of single-trajectory properties of fGn-driven non-Markovian processes through path-integral representations has recently received substantial attention. Path-integral methods offer a general formalism for stochastic dynamics, linking probabilistic descriptions with techniques from statistical field theory and quantum mechanics \cite{hibbs,wiegel,kleinert,wio,leticia,bing,theo}. They provide alternative approaches to computing correlation functions and allow for a systematic treatment of memory effects, external perturbations, and dynamical constraints. Within this framework, path-integral formulations have been developed for the trajectories of tagged beads in Gaussian polymer chains \cite{gleb} --- a subdiffusive process with Hurst exponent 
$H=1/4$
--- as well as for several variants of \textit{unconstrained} fractional Brownian motion \cite{seb,Jumarie,calvo,ben,ben2}. This line of work led to the recent study \cite{bar2}, which used the general form from \cite{ben2} to derive a path-integral representation of the fOUP driven by stationary fGn in the sense of Mandelbrot and van Ness \cite{man}. This analysis has been also performed for the unusual regime with $H \in (-1/2,0)$  \cite{ben3}.

In this paper, to complete the picture, we derive a path-integral representation of the fOUP driven by nonstationary Riemann–Liouville fractional Gaussian noise as defined by L\'evy \cite{levy} -- an alternative and widely used construction of fGn on a finite time $t$ interval $t \in (0,T)$, which is thus more suitable to investigate optimal problems on a finite time interval as compared to the Mandelbrot and van Ness fGn in which $t$ is defined on the entire real line~\cite{man}. We obtain two equivalent formulations of the action: one expressed explicitly in terms of fractional integrals \cite{samko}, and another written as a double integral with a nonlocal kernel. In addition, building on the general framework developed in \cite{baruch}, we derive explicit closed-form expressions for the optimal (action-minimizing) trajectories of the fOUP conditioned to reach a prescribed point 
$X >0$
at time 
$T'$, and analyze their qualitative behaviors. In particular, we show that in the subdiffusive regime, and for sufficiently strong confining potential, the optimal paths display a counterintuitive behavior: rather than approaching the target monotonically, they initially move away from 
$X$, then reverse the direction and reach the target in a rapid final excursion.

The paper is structured as follows: In Sec.~\ref{model} we formulate our model and introduce basic notations. In Sec. \ref{path-integrals} we present the derivation 
of the action in two alternative forms: in terms of fractional integrals and as a double integral with a non-local kernel.  In Sec.~\ref{paths} we discuss the behavior of the optimal paths. Finally, in Sec.~\ref{conc} we conclude with a brief recapitulation of our results. For completeness, the derivation of the covariance function of the Riemann-Liouville fOUP is given in \ref{A}.

\section{Fractional Ornstein-Uhlenbeck process with Riemann-Liouville fractional Gaussian noise}
\label{model}

Consider a stochastic differential equation of the form
\begin{align}
\label{a}
\gamma \dot{x}_t = -  \kappa x_t + \xi_t \,, \quad x_{t=0} = \dot{x}_{t=0} = 0 \,,
\end{align}
where the dot denotes the time derivative and $\xi_t$ is the Riemann-Liouville fractional Gaussian noise (RL fGn), defined by
\begin{align}
\label{1}
\xi_t = \dot{R}^{(H)}_t = \frac{1}{\Gamma(H+1/2)} \frac{d}{d t} \int^t_0 \frac{\zeta_{\tau}\, d\tau}{(t - \tau)^{ 1/2 - H}} \,,
\end{align}
where $H \in (0,1)$ is the Hurst index, $R^{(H)}_t$ is a given trajectory of the Riemann-Liouville fractional Brownian motion (RL fBm), first introduced in \cite{levy} (see also \cite{lim} and references therein), and $\zeta_{\tau}$ is a Gaussian white noise with zero mean and the covariance function
\begin{align}
\label{0}
\overline{\zeta_{\tau} \zeta_{\tau'}} = 2 D \delta(\tau - \tau') \,.
\end{align}
In the latter expression and henceforth the bar denotes averaging over realizations of 
the Gaussian noise and $D$ is its intensity. We choose throughout the unit of time such that $\gamma=1$.

The stochastic process defined in eq.~\eqref{a}, driven by the non-stationary fGn of eq.~\eqref{1}, constitutes the primary focus of our study, for which we aim to derive an exact path-integral representation of its individual trajectories 
$x_t$.
We note that the statistical properties of the corresponding process driven by \textit{stationary} fGn, as defined by Mandelbrot and van Ness \cite{man}, have been extensively investigated in both mathematical \cite{cheredito,salminen} and physical contexts (see, e.g., \cite{jeon,chev,ergo}). The path-integral representation for the fOUP driven by the Mandelbrot and van Ness fGn was recently obtained in~\cite{bar2}.

The fOUP $x_t$ in eq.~\eqref{a} is a zero-mean Gaussian stochastic process and its statistical properties are entirely defined by its covariance function $\cov_x(t,t')$. To determine this property, consider first
the covariance function of the RL fBm $R^{(H)}_t$ defined in 
eq.~\eqref{1}, which obeys, for arbitrary $t$, $t'$ and $H$, 
\begin{align}
\begin{split}
\label{cov_R}
\cov_{R}(t,t') & = \overline{R^{(H)}_t R^{(H)}_{t'}}\\ 
& = \frac{2 D}{\Gamma^2(H+1/2)} \int^m_0 \frac{d\tau}{\left[(t- \tau) (t' - \tau)\right]^{1/2 - H}} \,, \quad m = \min(t,t')  \,.
\end{split}
\end{align}
The above integral can be performed for arbitrary $t$, $t'$ and $H$ to give (see, e. g., \cite{seb})
\begin{equation}
	\label{covR}
	\begin{split}
	\cov_{R}(t,t') & = \frac{2 D m^{H+ 1/2} M^{H - 1/2}}{(H+1/2) \Gamma^2(H + 1/2)} \\
	&\qquad\qquad \times {}_2F_1\left(1/2-H, 1; H+ 3/2; \frac{m}{M}\right) \,, \quad M = \max(t,t') \,,
	\end{split}
	\end{equation}
where ${}_2F_1$ is Gauss's hypergeometric function.
Setting $t'=t$, one finds that the mean-square displacement of the RL fBm  $R^{(H)}_t$ obeys, for any $t$, 
\begin{align}
	\overline{\left(R^{(H)}_t\right)^2} = \frac{D}{H\Gamma^2(H+1/2)} \, t^{2H} \,.
	\end{align}
For 
$H > 1/2$, the dynamics of the process is superdiffusive, characterized by persistent temporal correlations whereby successive increments tend to maintain the same sign. In this regime, the mean-square displacement grows faster than linearly with time. The special case 
$H = 1/2$
corresponds to standard Brownian motion, for which correlations vanish and the mean-square displacement increases linearly. In contrast, for 
$H < 1/2$, the RL fBm 
exhibits subdiffusive or antipersistent behavior, characterized by \textit{negatively} correlated increments and a slower-than-linear growth of the mean-square displacement.

Note then that the covariance function of the RL fGn in eq.~\eqref{1} formally obeys
\begin{align}
	\label{7}
	\cov_{\xi}(t,t') = \frac{\partial^2}{\partial t\, \partial t'} \cov_{R}(t,t')  \,,
	\end{align}
and is a \textit{bona fide} function only for $H \geq 1/2$. For $H = 1/2$ the covariance is the delta-function, as it should be, while for  $H < 1/2$ the situation is quite delicate: here, the second derivative in eq.~\eqref{7} produces non-integrable diagonal singularities and must be interpreted only distributionally (or as a bilinear form on test functions). However, integrating that distribution against smooth kernels regularizes it and yields finite covariances of positions $x_t$ of the process in eq.~\eqref{a} (see, e.g., \cite{pipiras}).

\begin{figure}[ht]
\centering
\includegraphics[width=75mm]{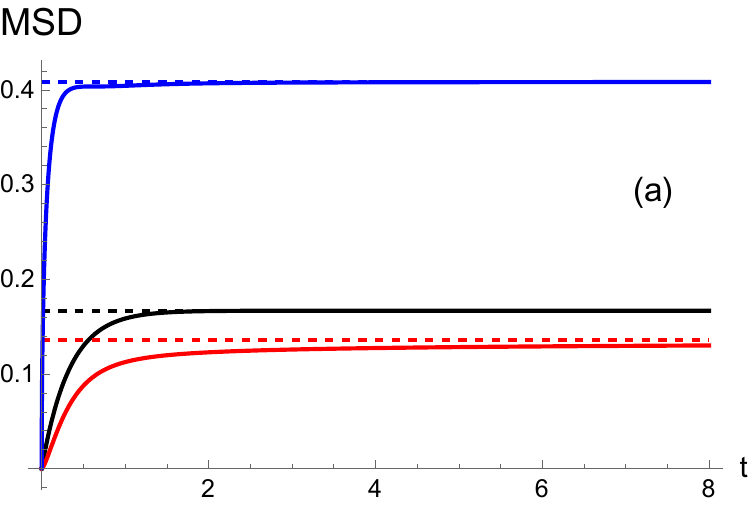}  
\includegraphics[width=75mm]{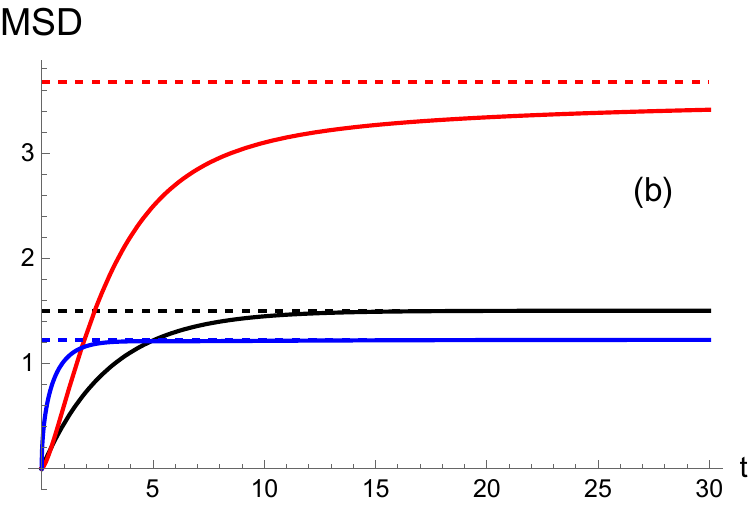}
\includegraphics[width=75mm]{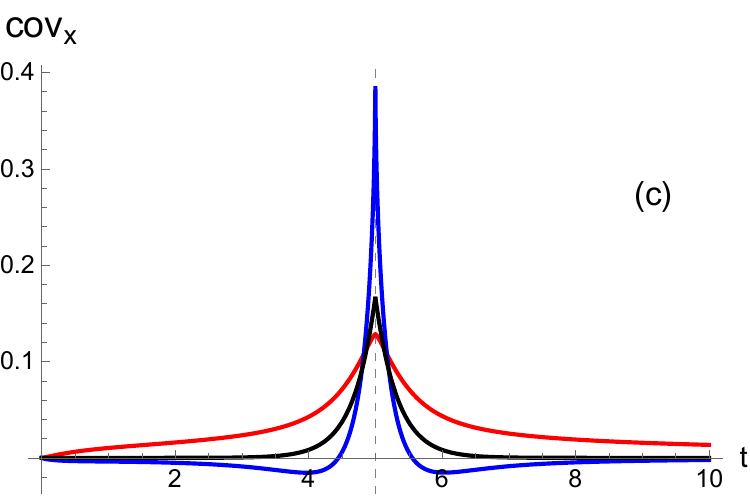}
\includegraphics[width=75mm]{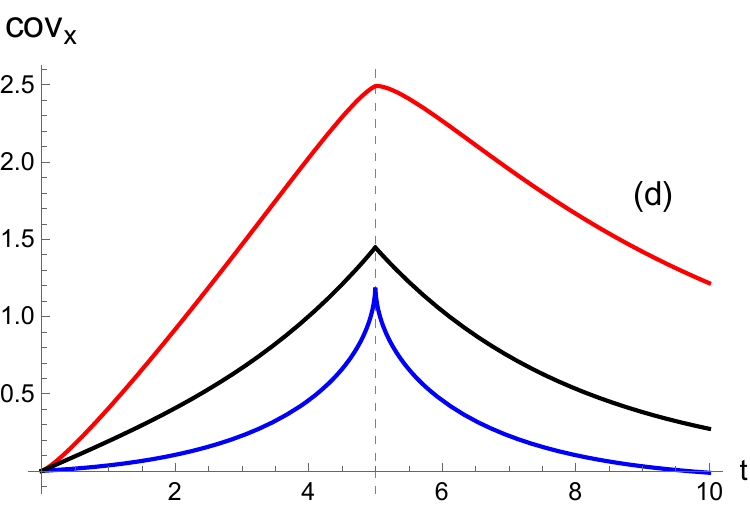}
\caption{Mean-square displacement (MSD) and the
	covariance function as functions of $t$ of the Ornstein-Uhlenbeck process driven by the Riemann-Liouville fractional Gaussian noise. Red curves depict the behavior in the superdiffusive case with $H = 3/4$, while the blue ones - in the subdiffusive case with $H = 1/4$.
	Black solid curves present the behavior for the standard Ornstein-Uhlenbeck process with $H = 1/2$. Units are such that $D =1/2$ and $\gamma=1$.
	Top row: The MSD, eq.~\eqref{MSD}, for $\tau^* = 1/3$  (panel (a)) and $\tau^* = 3$ (panel (b)). Bottom row: $\cov_x(t,t')$, eq.~\eqref{COVX}, 
	with $\tau^* = 1/3$ (panel (c)) and $\tau^* = 3$ (panel (d)). $t' = 5$ (vertical dashed line).}
\label{fig:1} 
\end{figure}
Respectively, the covariance function of the process $x_t$ is given by
\begin{align}
\label{covx}
\cov_x(t,t') = \overline{x_t x_{t'}} = \frac{1}{\gamma^2} e^{- (t + t')/\tau^*} \int^t_0 d\tau_1 \; e^{\tau_1/\tau^*} \int^{t'}_0 d\tau_2 \; e^{\tau_2/\tau^*} \, \cov_{\xi}(t,t')  \,,
\end{align}
where $\tau^*$ is the natural time scale,
$\tau^* = \gamma/\kappa$, where $\gamma=1$ in our units. Using eq.~\eqref{7} and the Beta-integral-type definition in eq.~\eqref{cov_R}, expression \eqref{covx} can be cast into a simpler form (see \ref{A}), 
\begin{align}
	\label{COVX}
\cov_x(t,t') =  \frac{2 D}{\Gamma^2(H+1/2)\gamma^2} \int^m_0 d\tau\; \frac{{}_1F_1\left(1, H+1/2, - \frac{t - \tau}{\tau^*}\right) {}_1F_1\left(1, H+1/2, - \frac{t' - \tau}{\tau^*}\right)}{\left[(t- \tau) (t' - \tau)\right]^{1/2 - H}}	\,,
	\end{align}
where ${}_1F_1$ is the confluent hypergeometric function. The expression \eqref{COVX} is formally valid for any $t$, $t'$ and $H$ and 
is more amenable to analytical and numerical analyses than the one in eq.~\eqref{covx}. Correspondingly, the MSD is given by
\begin{align}
	\label{MSD}
	\overline{x_t^2} = \frac{2 D}{\Gamma^2(H+1/2)\gamma^2} \int^t_0 d\tau\; \frac{\,{}_1F^2_1\left(1, H+1/2, - \frac{t - \tau}{\tau^*}\right)}{(t- \tau)^{1 - 2 H}} \,.
	\end{align}
Alternative representations of the covariance function and the MSD in terms of the Mittag-Leffler function are presented in \ref{A}.	
	
The integral in eq.~\eqref{MSD} cannot be performed in an explicit form but its asymptotic behavior in the limit $t \to \infty$ can  be readily determined to give
\begin{align}
	\label{MSD2}
	\overline{x_t^2} = \frac{  (\tau^*)^{2H}}{\sin(\pi H)} \frac{D}{\gamma^2} + O\left(e^{-t/\tau^*}\right) \,.
	\end{align}
This expression shows that the MSD relaxes exponentially in time to a finite asymptotic value that depends nontrivially on the time-scale 
$\tau^*$; namely, for $\tau^* > 1$, (i.e., $\gamma > \kappa$), the limiting MSD increases with the Hurst exponent 
$H$, attaining larger values in the superdiffusive regime than in the subdiffusive one. In contrast, 
 for $\tau^* < 1$, (i.e., $\gamma < \kappa$), the asymptotic MSD decreases with 
 $H$, so that subdiffusive dynamics yield larger stationary fluctuations than superdiffusive ones. A qualitatively similar dependence has been reported for the fOUP driven by the stationary fractional Gaussian noise introduced by Mandelbrot and van Ness (see, e.g., \cite{jeon}).
 
Figure \ref{fig:1} shows the mean-square displacement  and the covariance function as functions of time for two values of the characteristic time scale, 
$\tau^*=1/3$ and 
$\tau^* = 3$.
For each case, we compare a subdiffusive process with 
$H=1/4$
(blue curves) and a superdiffusive one with 
$H=3/4$
 (red curves), along with the classical OU process corresponding to 
$H=1/2$
(black curves). The MSD clearly illustrates the trend described above: when 
$\tau^* < 1$, the  MSD is larger in the subdiffusive regime than in the superdiffusive one, whereas for 
$\tau^* > 1$ the ordering is reversed.
The behavior of the covariance function is more nuanced. For 
$\tau^* < 1$, the same ordering holds only when 
$t$
and 
$t'$
are close to one another (the region that determines the MSD), while for sufficiently different $t$ and $t'$ the covariance in the subdiffusive case becomes smaller than in the superdiffusive case. For 
$\tau^* > 1$, by contrast, the covariance function corresponding to the superdiffusive process remains larger for all times $t$ and $t'$.
A notable feature in the subdiffusive case, particularly visible in Fig.~\ref{fig:1}(a), is the appearance of negative lobes in the covariance function away from the diagonal 
$t = t'$. This sign change is not an artefact but a direct manifestation of anti-persistence in the driving fGn: correlations at different times tend to oppose each other, and the fOUP inherits this behavior. Similar negative covariance regions are also known for fOUP driven by the stationary fractional Gaussian noise of Mandelbrot and van Ness (see, e.g., \cite{bar2}).

\section{Path-integral representations of the fractional Ornstein-Uhlenbeck process}
\label{path-integrals}  
Using the definition in eq.~\eqref{1} we rewrite formally eq.~\eqref{a} as
\begin{align}
 \frac{1}{\Gamma(H+1/2)}  \int^t_0 \frac{\zeta_{\tau}\, d\tau}{(t - \tau)^{1/2-H}} = K_t \,, \quad K_t = \int^t_0 d\tau \left(\gamma \dot{x}_{\tau} + \kappa x_{\tau}\right).
\end{align}
This is Abel's integral equation which can be solved exactly by applying to both sides of this equation (from the left) an appropriate inverse Riemann-Liouville operator (see \cite[p.~29]{samko}), which gives
\begin{align}
\label{b}
\zeta_t = \frac{1}{\Gamma(1/2-H)} \frac{d}{d t} \int^t_0 \frac{K_{\tau}\, d\tau}{(t - \tau)^{H+1/2}} = \left({\cal D}^{H+1/2}_{+}K \right)(t) \,,
\end{align} 
where the operator ${\cal D}^{H+1/2}_{+} K$ in the right-hand-side (rhs) of eq.~\eqref{b} denotes the so-called \textit{fractional} Riemann-Liouville derivative of order $H + 1/2$ acting on the function $K$ (see the definition 2.2  in \cite[p.~35]{samko}). The subscript $+$ indicates that we have the ``left-handed" derivative. 

For subdiffusive (antipersistent) noise the order $H + 1/2$  of the fractional derivative is less than $1$. In this case, we have (see \cite[p.~25, eq.~(2.24)]{samko})
\begin{align}
\label{c}
\left({\cal D}^{H+1/2}_{+}K \right)(t)  = \frac{1}{\Gamma(1/2-H)} \left[\frac{K_0}{t^{H+1/2}} + \int^t_0 \frac{\dot{K}_{\tau}\, d\tau}{(t - \tau)^{H+1/2}}\right] \,,
\end{align}
where the first term in the brackets evidently vanishes because $K_{t = 0} \equiv 0$. Consequently, 
for subdiffusive fractional Riemann-Liouville noise one has
\begin{align}
\label{z1}
\zeta_t = \left({\cal D}^{H+1/2}_{+}K \right)(t)  = \frac{1}{\Gamma(1/2-H)}  \int^t_0 \frac{\left(\gamma \dot{x}_{\tau} + \kappa x_{\tau}\right)\, d\tau}{(t - \tau)^{H+1/2}} \,.
\end{align}

For superdiffusive (persistent) noise the order $H + 1/2$ of the fractional derivative  is greater than $1$ and can be written as 
$H + 1/2 = 1 + (H - 1/2)$. We take advantage of the representation in \cite[eq.~(2.43)]{samko} to get
\begin{align}
	\begin{split}
\left({\cal D}^{H+1/2}_{+}K \right)(t)  &= \frac{K_0}{\Gamma(1/2-H) t^{H+1/2}} + \frac{\dot{K}_0}{\Gamma(3/2-H) t^{H - 1/2}}\\ &\qquad\qquad {}+ \frac{1}{\Gamma(3/2 - H)} \int^{t}_0 \frac{\ddot{K}_{\tau}\, d\tau}{(t - \tau)^{H - 1/2}} \,.
\end{split}
\end{align}
The first term in the rhs of the latter expression is identically equal to zero, while the second one vanishes because of the initial conditions we have chosen. Consequently,  for the superdiffusive fractional Riemann-Liouville noise we get
\begin{align}
\label{z2}
\zeta_t = \left({\cal D}^{H+1/2}_{+}K \right)(t)  = \frac{1}{\Gamma(3/2 - H)} \int^{t}_0 \frac{\left(\gamma \ddot{x}_{\tau} + \kappa \dot{x}_{\tau}\right) d\tau}{(t - \tau)^{H - 1/2}} \,.
\end{align}

Lastly, the probability functional for observing a given realization of Gaussian white noise $\zeta_t$ on a finite time interval $(0,T)$ is
\begin{align}
	\label{ddd}
P[\zeta_t] \simeq \exp\left( - \frac{1}{4 D} \int^T_0 dt \, \zeta_t^2 \right) \,, 
\end{align}
which is central for our further analysis of the form of the probability $P[x_t]$ of observing a given trajectory $x_t$ of the fOUP on the interval $(0,T)$. Capitalizing on eq.~\eqref{ddd} and eqs.~\eqref{z1} and \eqref{z2}, we derive below the desired path-integrals representations of fOUP.

\subsection{Path-integral representations in terms of fractional integrals}

We seek first an exact   
representation of $P[x_t]$ with an action written 
in terms of fractional integrals, which can be done very straightforwardly by merely using
eq.~\eqref{b}. We find then that the probability of observing on the time-interval $(0,T)$ a given realization $x_t$ of the process obeying eq.~\eqref{a}
 is formally given by
\begin{align}
P[x_t] \simeq \exp\left( - \frac{1}{4 D} \int^T_0 dt \left[ \left({\cal D}^{H+1/2}_{+}K \right)(t) \right]^2 \right) \,.
\end{align}
Further on, taking advantage of expressions \eqref{z1} and \eqref{z2},
 we find
\begin{align}
\begin{split}
\label{d}
P[x_t] &\simeq \exp\left( - \frac{1}{4 D} \int^T_0 dt \left[ \frac{1}{\Gamma(1/2-H)}  \int^t_0 \frac{\left(\gamma \dot{x}_{\tau} + \kappa x_{\tau}\right) d\tau}{(t - \tau)^{H+1/2}}\right]^2 \right) \,,\\
P[x_t] &\simeq \exp\left( - \frac{1}{4 D} \int^T_0 dt \left[\frac{1}{\Gamma(3/2 - H)} \int^{t}_0 \frac{\left(\gamma \ddot{x}_{\tau} + \kappa \dot{x}_{\tau}\right) d\tau}{(t - \tau)^{H - 1/2}} \right]^2 \right)  \,,
\end{split}
\end{align}
for the subdiffusive and superdiffusive cases, respectively.

Three remarks regarding the expressions in eqs. \eqref{d} are in order.
First, setting 
$\kappa = 0$
one recovers known results for unconstrained dynamics. In this limit, the first line of eq.~\eqref{d} reduces to the classical action for the Riemann-Liouville fractional Brownian motion obtained in \cite{seb}, while the second line coincides with the superdiffusive result reported in \cite{ben}.
Second, the appearance of second-order derivatives in the action for the superdiffusive regime is expected: such terms arise naturally in systems with bending rigidity, such as semiflexible polymers and membranes \cite{bing}, for random acceleration process \cite{theo}, and for unconstrained superdiffusive fractional Brownian motions \cite{ben,ben2}.  
Third, we consider the limit $H \to 1/2$, i.e., the case of a white noise. In the superdiffusive case (second line in eq.~\eqref{d}) we can straightforwardly set $H = 1/2$ in the fractional integral to get
\begin{align}
\begin{split}
\label{OU}
P[x_t] &\simeq \exp\left( - \frac{1}{4 D} \int^T_0 dt \left[ \int^{t}_0 \left(\gamma \ddot{x}_{\tau} + \kappa \dot{x}_{\tau}\right) d\tau \right]^2 \right)  \\
&\simeq \exp\left( - \frac{1}{4 D} \int^T_0 dt \left(\gamma \dot{x}_{t} + \kappa x_{t}\right)^2 \right) \,,
\end{split}
\end{align}
which is the exact result for the standard OUP driven by a Gaussian white noise. In the subdiffusive case the passage $H \to 1/2$ is a bit more delicate and cannot be performed directly in the integral. To this end, we first formally rewrite the fractional integral as
\begin{align}
\frac{1}{\Gamma(1/2-H)}  \int^t_0 \frac{\left(\gamma \dot{x}_{\tau} + \kappa x_{\tau}\right) d\tau}{(t - \tau)^{H+1/2}} = - \frac{1}{\Gamma(3/2-H)}  \int^t_0 \left(\gamma \dot{x}_{\tau} + \kappa x_{\tau}\right) d\tau \frac{d}{d \tau} (t - \tau)^{1/2 - H}\,,
\end{align}
and then perform the integral in the right-hand-side by parts assuming that $H < 1/2$, i.e., $H$ is bounded away from $1/2$. This gives
\begin{equation}
\begin{split}
&- \frac{1}{\Gamma(3/2-H)}  \int^t_0 \left(\gamma \dot{x}_{\tau} + \kappa x_{\tau}\right) d\tau \frac{d}{d \tau} (t - \tau)^{1/2 - H}\\
&\qquad\qquad {}=  \frac{1}{\Gamma(3/2-H)}  \int^t_0 \left(\gamma \ddot{x}_{\tau} + \kappa \dot{x}_{\tau}\right) (t - \tau)^{1/2 - H} d\tau \,,
\end{split}
\end{equation}
in which we can now safely put $H = 1/2$ to get the expression \eqref{OU}.

\subsection{Path-integral representations with non-local kernels}

We now focus on the path-integral representation in which the action takes a more standard form of the double integral with a non-local kernel. Changing the integration order in eq.~\eqref{d} and performing the inner integral over $t$,  we have that in case of a subdiffusive fOUP $P[x_t]$ attains the form
\begin{equation}
	\begin{split}
		\label{d_sub}
		P[x_t] &\simeq \exp\left( - \frac{1}{4 D \Gamma^2(1/2-H)} \int_{0}^T d\tau_1 \left(\gamma \dot{x}_{\tau_1} + \kappa x_{\tau_1}\right)\right.\\
		 &\qquad\qquad \left. {}\times\int_{0}^T d\tau_2 \left(\gamma \dot{x}_{\tau_2} + \kappa x_{\tau_2}\right) \, q_{\sub}(\tau_1,\tau_2)\right)  \,,
	\end{split}
\end{equation}
where the kernel $q_{\sub}(\tau_1,\tau_2)$ is a piece-wise continuous function of $\tau_1$ and $\tau_2$ which is given explicitly by
\begin{align}
	\label{Sub}
	q_{\sub}(\tau_1,\tau_2) = \frac{1}{|\tau_1 - \tau_2|^{2 H}} B_z\left(1/2-H,2 H\right) \,, \quad z = \frac{T - \max(\tau_1,\tau_2)}{T - \min(\tau_1,\tau_2)}\,,
	\end{align}
where $B_z(a,b)$ is the incomplete Beta function:
\begin{align}
B_z(a,b) = \int^z_0 dx \; x^{a - 1} (1 - x)^{b-1} \,.	
	\end{align}
We depict $q_{\sub}$ as function of $\tau_2$ (with fixed $\tau_1$) in the left panel in Fig.~\ref{fig:3}. We observe that  $q_{\sub}$ tends to a computable constant when $\tau_2  \to 0$, vanishes when $\tau_2 \to T$ and diverges when $\tau_2 \to \tau_1$,
\begin{align}
	\label{subb}
q_{\sub}(\tau_1,\tau_2) \simeq \frac{\Gamma(1/2-H) \Gamma(2 H)}{\Gamma(1/2 + H)} \frac{1}{|\tau_1 - \tau_2|^{2 H}}	\,.
	\end{align}
The divergence is stronger the closer $H$ is to $1/2$.

Consider next the case of superdiffusion. Performing essentially the same procedure, we get
\begin{align}
	\begin{split}
		\label{d_sup}
		P[x_t] &\simeq \exp\left( - \frac{1}{4 D \Gamma^2(3/2-H)} \int_{0}^T d\tau_1 \left(\gamma \ddot{x}_{\tau_1} + \kappa \dot{x}_{\tau_1}\right)\right.\\
		&\qquad\qquad \left. {}\times \int_{0}^T d\tau_2 \left(\gamma \ddot{x}_{\tau_2} + \kappa \dot{x}_{\tau_2}\right) \, q_{\Sup}(\tau_1,\tau_2)
		\right)  \,,
	\end{split}
\end{align}
where the kernel $q_{\Sup}(\tau_1,\tau_2)$ obeys
\begin{align}
	\label{Sup}
	q_{\Sup}(\tau_1,\tau_2) = |\tau_1 - \tau_2|^{2 - 2H} B_z\left(3/2-H,2 H - 2\right) \,.
\end{align}
This kernel is depicted as function of $\tau_2$ (for fixed $\tau_1$) in Fig.~\ref{fig:3}. Likewise in the subdiffusive case, $q_{\Sup}$ tends to a computable constant when $\tau_2 \to 0$ and vanishes when $\tau_2 \to T$. In contrast to the subdiffusive case, $q_{\Sup}$ approaches a constant value when $\tau_2 \to \tau_1$. Indeed, the Beta function in eq.~\eqref{Sup} diverges as 
\begin{align}
	B_z\left(3/2-H,2 H - 2\right) \simeq \frac{(3 - 2 H)}{4(1-H)} (1 - z)^{2H - 2} \,, \quad z \to 1 \,,
	\end{align}
and consequently,
\begin{align}
	\lim_{\tau_2 \to \tau_1} q_{\Sup}(\tau_1,\tau_2) =
\frac{(3 - 2 H)}{4(1-H)} (T - \tau_1)^{2 (1 - H)} \,.	 
	\end{align} 
Therefore, for $\tau_2 = \tau_1$ and finite $T$ the kernel has a finite-height peak when the arguments coincide. The height of the peak depends on $\tau_1$ and $T$.

\begin{figure}[ht]
	\centering
	\includegraphics[width=75mm]{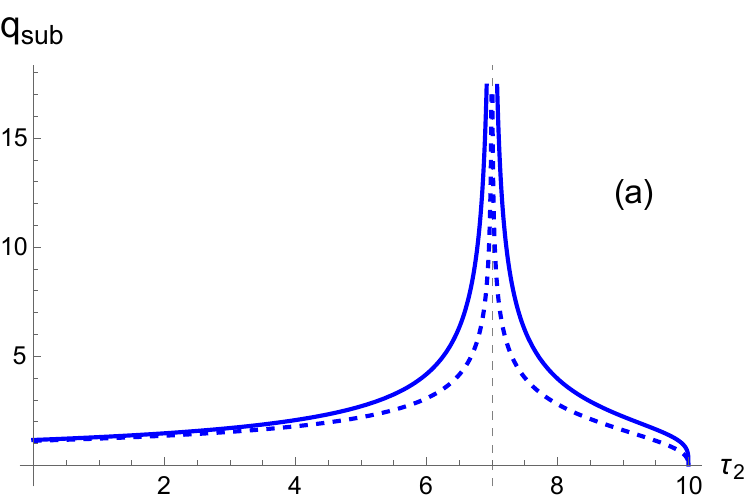}  
	\includegraphics[width=75mm]{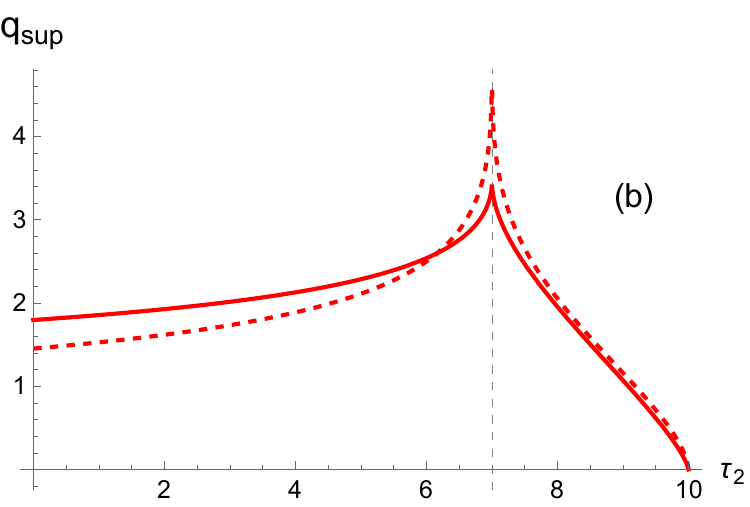}
	\caption{Kernels $q_{\sub}(\tau_1,\tau_2)$ and $q_{\Sup}(\tau_1,\tau_2)$ in eqs.\eqref{Sub} and \eqref{Sup} as functions of $\tau_2$ for $\tau_1 = 7$ (vertical dashed line) and $T = 10$.  
		Panel (a): Subdiffusive case.  The kernel $q_{\sub}(\tau_1,\tau_2)$ in eq.~\eqref{Sub} with $H = 1/4$ (blue solid curve) and $H = 1/8$ (blue dashed curve). Panel (b): Superdiffusive case. The kernel $q_{\Sup}(\tau_1,\tau_2)$ in eq.\eqref{Sup}  for $H=7/8$ (red dashed curve) and $H = 3/4$ (red solid curve).}
	\label{fig:3} 
\end{figure}

\section{Optimal paths}
\label{paths}

Consider such paths $x^*_t$ which start at $0$ at $t=0$, appear at position $X$ at time instant $T' < T$, and provide the minimum value to the actions in eqs. \eqref{d_sub} or \eqref{d_sup}; that being, $x^*_t$ are the action-minimizing or ``optimal" paths. 
Our aim is to define $x^*_t$ explicitly; to this end, we resort to the analysis  in recent \cite{baruch} in which this problem has been solved for arbitrary Gaussian processes which are entirely defined by their covariance function. For completeness, we repeat here the main arguments presented in \cite{baruch}.

\begin{figure}[ht]
	\centering
	\includegraphics[width=75mm]{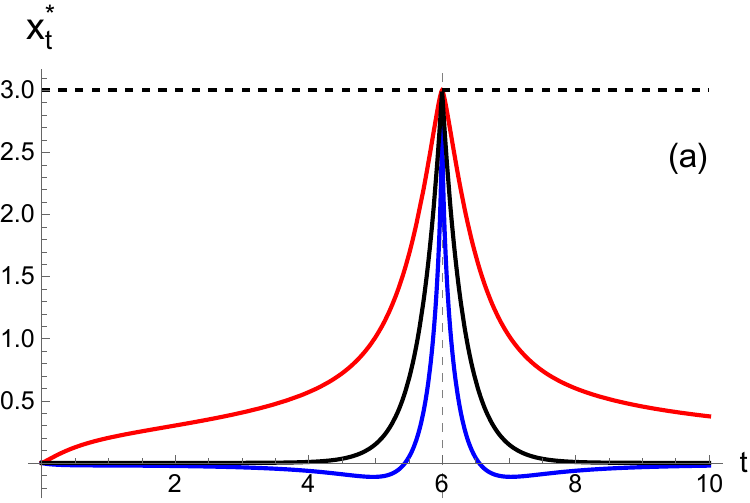}  
	\includegraphics[width=75mm]{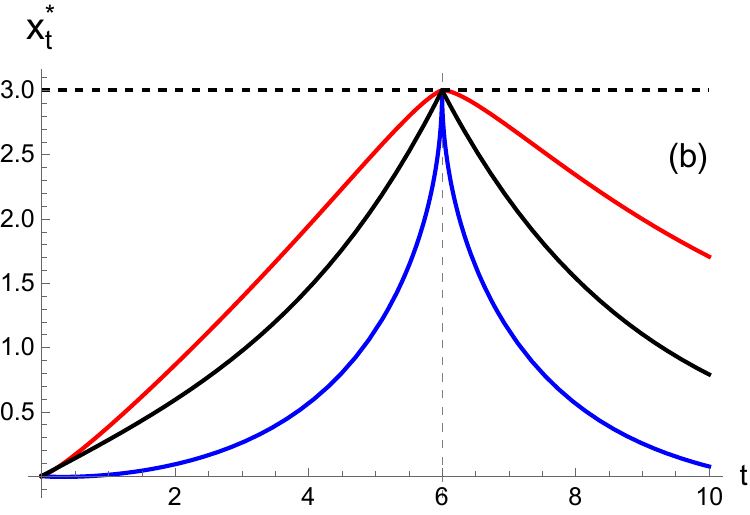}
	\includegraphics[width=75mm]{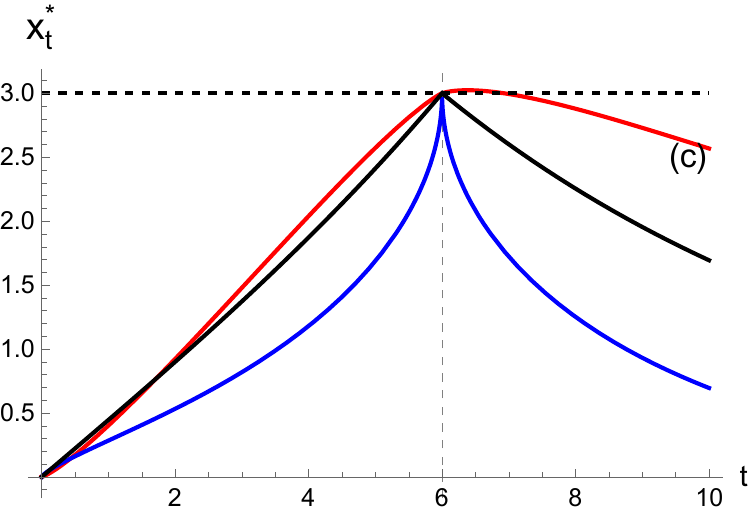}
	\includegraphics[width=75mm]{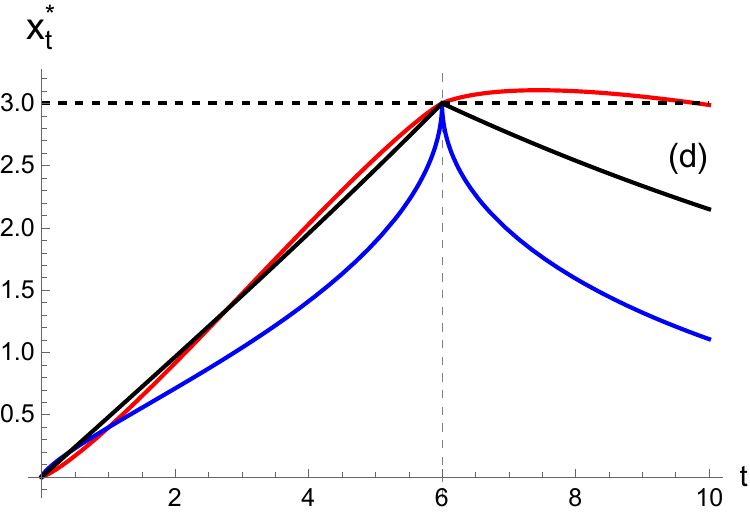}
	\caption{Optimal paths $x^*_t$ with $t \in (0,T)$ (here, $T = 10$) conditioned to be at point $X$ (here, $X = 3$) at time moment $T'$ (here, $T'=6$, vertical dashed line) for superdiffusive regime with $H = 3/4$ (red curves) and subdiffusive regime with $H = 1/4$ (blue curves). Black solid curves depict the corresponding optimal paths for the standard OU process ($H = 1/2$). In four panels we present $x_t^*$ for different values of $\tau^*$: 
		$\tau^* = 1/3$ (panel (a)), $\tau^* = 3$ (panel (b)), $\tau^* = 7$ (panel (c)) and $\tau^*=12$ (panel(d)).}
	\label{fig:3} 
\end{figure}

Suppose that we rewrite formally, by integration in parts, the actions in eqs. \eqref{d_sub} and \eqref{d_sup} in the canonical form 
\begin{align}
	S = \frac{1}{4 D} \int^T_0 d\tau_1  \int^T_0 d\tau_2 \, x_{\tau_1} \, x_{\tau_2} \, Q(\tau_1,\tau_2) \,,
	\end{align} 
where $Q(\tau_1,\tau_2) = Q(\tau_2,\tau_1)$ can be readily expressed in terms of $q_{\sub}(\tau_1,\tau_2)$ in case of subdiffusion or $q_{\Sup}(\tau_1,\tau_2)$ for the superdiffusive motion. The resulting expressions are quite lengthy and we do not present them here. As a matter of fact, one does not need to know the precise form of this kernel. On the other hand, the kernel $Q(\tau_1,\tau_2)$ obeys the integral equation \cite{zinn}
\begin{align}
	\label{z9}
	\int^T_0 d\tau_2 \, Q(\tau_1,\tau_2) \, \cov_x(\tau_1',\tau_2) = \delta(\tau_1 - \tau_1') \,,
	\end{align}
where $\cov_x(\tau_1',\tau_2)$ is the covariance function of the process $x_t$ defined in eq.~\eqref{COVX}. Next, we consider an auxiliary action of the form 
\begin{align}
	\tilde{S} = \frac{1}{4 D} \int^T_0 d\tau_1 \left[\int^T_0 d\tau_2 \, x_{\tau_1} \, x_{\tau_2} \, Q(\tau_1, \tau_2) - \lambda \, x_{\tau_1} \delta(\tau_1 - T') \right] \,,
	\end{align}
where the second term in the brackets implements the condition
\begin{align}
	\int^T_0 d\tau_1 \, x_{\tau_1} \, \delta(\tau_1 - T') = 
	X \,, 
\end{align}
while the Lagrange multiplier $\lambda$ is to be choosen to ensure that $x_{t = T'} = X$. Then, the linear variation of the auxiliary action is given by
\begin{align}
		\delta\tilde{S} = \frac{1}{2 D} \int^T_0 d\tau_1 \delta x_{\tau_1}\left[\int^T_0 d\tau_2  \, x_{\tau_2} \, Q(\tau_1, \tau_2) - \frac{\lambda}{2} \, \delta(\tau_1 - T') \right] \,,
	\end{align}
and must vanish for arbitrary $\delta x_{\tau_1}$ which yields a linear integral equation for the optimal paths:
\begin{align}
	\label{z10} 
	\int^T_0 d\tau_2  \, x^*_{\tau_2} \, Q(\tau_1, \tau_2) = \frac{\lambda}{2} \, \delta(\tau_1 - T')
	\end{align}
Comparing eqs.~\eqref{z10} and \eqref{z9} one readily infers that \cite{baruch}
\begin{align}
	x^*_t = \frac{\lambda}{2} {\rm cov_x}(t,T')
	\end{align}
and hence, chosing $\lambda = 2  X/{\rm cov_x}(T',T')$, one has that the optimal path from the origin to point $X$ reached at time instant $t = T' < T$ obeys
\begin{align}
	\label{opt}
x^*_t = \frac{{\rm cov_x}(t,T')}{{\rm cov_x}(T',T')} X \,.	
	\end{align} 
Note that the above procedure can be readily generalized 
for the optimal paths constrained to visit point $X_1$ at time moment $T_1$, point $X_2$ at time moment $T_2$, and so on (see \cite{baruch}).
 
Figure \ref{fig:3} shows the optimal paths 
$x_t^*$
from eq. \eqref{opt}, conditioned to start at the origin at 
$t=0$
and to reach a prescribed point 
$X >0$
at time 
$t = T'$. The red curves correspond to the superdiffusive regime ($H=3/4$), the blue curves to the subdiffusive regime ($H=1/4$), and the black curves represent the classical OUP case ($H=1/2$). The comparison highlights how persistence or anti-persistence of the driving noise affects the structure of the optimal trajectories and makes it different from the OUP case.
The four panels of Fig. \ref{fig:3} correspond to different values of the parameter  $\tau^* = \gamma/\kappa$: 
$\tau^* = 1/3$ (a), $\tau^* = 3$ (b), $\tau^* = 7$ (c)
and $\tau^* = 12$ (d). Panels (a) and (b) both correspond to $\tau^* < T' < T$, but differ in whether 
$\tau^*$ is smaller or larger than unity, leading to qualitatively distinct behaviors. In panel (c), the scales satisfy 
$T' < \tau^* < T$, while in panel (d) one has 
$T' < T < \tau^*$. Since 
$\tau^*$ is inversely proportional to the strength of the confining quadratic potential, panel (a) corresponds to the strongest confinement and panel (d) to the weakest.

We begin with the superdiffusive case. In all four panels, the optimal trajectories approach the target monotonically, steadily decreasing their distance to 
$X$. Under strong confinement (panel (a)), the velocity $\dot{x}^*_t$ varies noticeably in time, while this variation becomes progressively weaker as 
$\tau^*$ increases, corresponding to a reduction of the confining strength. One also observes that the optimal fOUP trajectory initially moves more slowly than the classical OUP, but then accelerates at later times (see panels (c) and (d)). After reaching the target, the behavior depends on the confinement strength: in panels (a) and (b), the optimal path subsequently turns back for $t \geq T'$ toward the origin, whereas in panels (c) and (d) it overshoots, moving further away for a while before eventually reversing direction.

The subdiffusive case under strong confinement displays a markedly nontrivial behavior. Owing to the anti-persistence of the driving noise, the optimal trajectory initially moves \textit{away} from the target, increasing its distance from 
$X$, before abruptly reversing direction, accelerating, and reaching 
$X$ within a short time interval. For 
$t \geq T'$, the trajectory departs from the target rapidly, crosses the origin, and then relaxes toward it from below. As the confining potential is weakened (i.e., as 
$\tau^*$ increases), this pronounced manifestation of noise anti-correlations gradually disappears: the optimal paths then approach 
$X$ monotonically and subsequently drift back toward the origin. We note in passing that essentially the same behavior is observed for the Mandelbrot and van Ness fGn considered in \cite{bar2}. As seen in the left panel in  Fig. 2 therein, the optimal paths initially move to negative values -- i.e., away from the target -- before crossing the origin and approaching 
$X$. This indicates that such a non-monotonic behavior of the optimal paths is a generic feature of the fOUP driven by fGn, independent of the specific definition of the noise.

\section{Conclusions}
\label{conc}

In conclusion, we have developed a path-integral formulation for the non-Markovian fractional Ornstein–Uhlenbeck process driven by nonstationary Riemann–Liouville fractional Gaussian noise, covering both subdiffusive and superdiffusive regimes. Two equivalent representations of the associated action were obtained. The first expresses the action explicitly in terms of Riemann–Liouville fractional integrals, thereby making the role of fractional operators in the dynamics transparent. The second recasts the action in a more conventional double-integral form, at the expense of introducing a nonlocal kernel that encapsulates the memory inherent in the driving noise.

Within this framework, we further derived closed-form expressions for the optimal (action-minimizing) trajectories of the process conditioned to reach a prescribed position 
$X>0$
at some fixed time instant. The resulting paths were analyzed in detail across several parameter regimes, in particular, for different strengths of the confining potential. Interestingly enough, we found that in the subdiffusive case, and for sufficiently strong confining potentials, the optimal conditioned trajectories exhibit a non-monotonic structure: instead of moving directly toward the target, they initially deviate in the opposite direction before reversing and approaching 
$X$
through a rapid final excursion. This transient “overshoot” away from the target appears to be a robust feature of memory-driven relaxation and highlights qualitative differences between fractional and Markovian Ornstein–Uhlenbeck dynamics.

Our results provide a systematic foundation for the analysis of conditioned paths in fractional stochastic processes and may prove useful in applications where non-Markovian noise and constrained dynamics play a central role, including viscoelastic transport, intracellular biophysics, and anomalous relaxation in complex media.

\section*{Acknowledgments}

The authors wish to thank Baruch Meerson for helpful discussions. BM acknowledges support from the National Natural Science
Foundation of China (NSFC) (Grant Nos. 12575045 and 12034019).

\appendix

\section{Covariance function of the Riemann-Liouville fractional Ornstein-Uhlenbeck process.}
\label{A}

We first formally rewrite the definition \eqref{cov_R} as
\begin{align}
	\begin{split}
		\cov_{R}(\tau_1,\tau_2) & =  \frac{2 D}{\Gamma^2(H+1/2)} \int^{\infty}_0 \frac{d\tau \theta(\tau_1 - \tau) \theta(\tau_2 - \tau)}{\left[(\tau_1- \tau) (\tau_2 - \tau)\right]^{1/2 - H}}  \,,
	\end{split}
\end{align}
where $\theta(t)$ is the Heaviside theta function, which satisfies $\theta(t) = 1$ for $t\geq 0$, and zero otherwise.
Inserting the above expression into eq.~\eqref{covx} and changing the integration order, we have
\begin{align}
	\begin{split}
		\cov_x(t,t') &= \frac{2 D}{\Gamma^2(H+1/2) \gamma^2} e^{-(t + t')/\tau^*} \int^{\infty}_0 d\tau \left\{\int^t_0 d\tau_1 \;e^{\tau_1/\tau^*} \frac{d}{d \tau_1} \frac{\theta(\tau_1 - \tau)}{(\tau_1 - \tau)^{1/2-H}}\right\}\\
		&\qquad\qquad \times \left\{\int^t_0 d\tau_2\; e^{\tau_2/\tau^*} \frac{d}{d \tau_2} \frac{\theta(\tau_2 - \tau)}{(\tau_2 - \tau)^{1/2-H}}\right\} \,.
		\end{split}
		\end{align}
Next, integrating by parts the terms in the curly brackets, we have
\begin{align}
	\begin{split}
		\label{AZ}
\int^t_0 d\tau_1 \, e^{\tau_1/\tau^*}& \frac{d}{d \tau_1} \frac{\theta(\tau_1 - \tau)}{(\tau_1 - \tau)^{1/2-H}}	= \frac{\theta(t - \tau)}{(t - \tau)^{1/2-H}} \, e^{t/\tau^*}\\&\times	\left(1 - \frac{(t - \tau)}{(H + 1/2)\tau^*} \,_1F_1\left(1, H+3/2, -\frac{(t - \tau)}{\tau^*}\right)\right) \,,\\
\int^{t'}_0 d\tau_2 \, e^{\tau_2/\tau^*}& \frac{d}{d \tau_2} \frac{\theta(\tau_2 - \tau)}{(\tau_2 - \tau)^{1/2-H}}	= \frac{\theta(t' - \tau)}{(t' - \tau)^{1/2-H}} \, e^{t'/\tau^*}\\&\times	\left(1 - \frac{(t' - \tau)}{(H + 1/2) \tau^*} \,_1F_1\left(1, H+3/2, -\frac{(t' - \tau)}{\tau^*}\right)\right) \,.
			\end{split}
		\end{align}		
Using the identity
\begin{align}
	\frac{z}{b} \,{}_1F_1(a+1,b+1,z) =   {}_1F_1(a+1,b,z) -  {}_1F_1(a,b,z) \,,
	\end{align}
and rewriting formally the expressions in the right-hand-side of eqs.~\eqref{AZ},
we recover our results in eqs.~\eqref{COVX} and \eqref{MSD}. 

Finally, we note that 
\begin{align}
	{}_1F_1\left(1, H+1/2, z\right) = \Gamma(H+1/2) \sum_{n=0}^{\infty} \frac{z^n}{\Gamma(n + H +1/2)} =\Gamma(H+1/2) \, E_{1,H+1/2}(z) \,, 
	\end{align}
where $E_{1,H+1/2}(z)$ is the 
Mittag-Leffler function. Correspondingly, the representations of the covariance function and the MSD in eqs. \eqref{COVX} and \eqref{MSD} can be formally rewritten in terms of $E_{1,H+1/2}(z)$ as
\begin{align}
	\label{COVXML}
	\cov_x(t,t') =  \frac{2 D}{\gamma^2} \int^m_0 d\tau \frac{E_{1,H+1/2}\left(- \frac{t - \tau}{\tau^*}\right) \, E_{1,H+1/2}\left( - \frac{t' - \tau}{\tau^*}\right)}{\left[(t- \tau) (t' - \tau)\right]^{1/2 - H}}	\,,\quad m =\min(t,t') \,,
\end{align}
and
\begin{align}
	\begin{split}
	\label{MSDML}
	\overline{x_t^2} &= \frac{2 D}{\gamma^2} \int^t_0 d\tau \frac{E^2_{1,H+1/2}\left(- \frac{t - \tau}{\tau^*}\right)}{(t- \tau)^{1 - 2 H}} 
	= \frac{2 D}{\gamma^2} (\tau^*)^{2H} \int^{t/\tau^*}_0 dz \, z^{2H - 1} \, E^2_{1,H+1/2}(-z) \,.
	\end{split}
\end{align}

\end{document}